\documentclass[
    reprint,
    preprintnumbers,
    superscriptaddress,
    nofootinbib,
     amsmath,amssymb,
     aps,
     prd,
    floatfix, longbibliography
    ]{revtex4-2}

    \usepackage{graphicx}
    \usepackage[caption=false]{subfig}
    \usepackage[usenames,dvipsnames]{xcolor} 
    \usepackage{pgfplots}
    \usepackage[utf8]{inputenc}
    \usepackage{color}
    \usepackage{hyperref}
    \usepackage[normalem]{ulem} 
    \usepackage{physics}
    \usepackage{enumitem}
    \usepackage{comment}
    \usepackage{bm}
    \usepackage{booktabs}
\hypersetup{
  breaklinks = true,
  colorlinks   = true, 
  urlcolor     = Blue, 
  linkcolor    = Blue, 
  citecolor   = Blue 
}
    \allowdisplaybreaks[1]
    \graphicspath{{figures/}}

\usepackage{float}

\newcommand{\oxford}{Astrophysics, University of Oxford, DWB, Keble Road, Oxford OX1 3RH, United Kingdom}

\newcommand{\splitatcommas}[1]{%
  \begingroup
  \begingroup\lccode`~=`, \lowercase{\endgroup
    \edef~{\mathchar\the\mathcode`, \penalty0 \noexpand\hspace{0pt plus 1em}}%
  }\mathcode`,="8000 #1%
  \endgroup
}

\begin{document}

\title{Inflationary attractors and radiative corrections in light of ACT data}

\author{William J. Wolf}
\email{william.wolf@stx.ox.ac.uk}
\affiliation{\oxford}

\begin{abstract}
In light of the recent results from the Atacama Cosmology Telescope (ACT), which have provided a notable shift in the constraints on $(n_s, r)$ and placed several otherwise viable models of inflation in tension with the latest data, we investigate the possible effects that radiative corrections can have on $\xi$-attractor and $\alpha$-attractor models of inflation. These models, which share much in common with Starobinsky inflation, have likewise been put under pressure by these results. We find that percent (and even sub-percent) level radiative corrections can easily shift both of these classes of inflation models comfortably into the regions of parameter space favoured by the most recent constraints. However, the flexibility under such corrections calls into question to what extent it is possible to precisely pin down model-specific predictions for important cosmological observables. 
\end{abstract}

\maketitle


\section{Introduction}\label{sec:introduction}

Cosmic inflation has widely been considered to be the best paradigm to model the early universe ever since its introduction \cite{Starobinsky:1980te, Guth:1980zm, Albrecht:1982wi, Bardeen:1983qw, Linde:1981mu, Mukhanov:1981xt, Hawking:1982cz, Linde:1983gd}.  
In this recent era of precision cosmology, we have accurately measured the statistical features of the primordial density fluctuations that seed cosmic structure and found that they are in remarkable agreement with generic inflationary predictions \cite{Planck:2018jri, Chowdhury:2019otk, Kallosh:2025ijd, Guth:2013sya}. While there are a plethora of modeling constructs \cite{Martin:2013tda}, the Starobinsky \cite{Starobinsky:1980te, Vilenkin:1985md, Ketov:2025nkr} and Higgs-like models of inflation \cite{Bezrukov:2007ep, Bezrukov:2010jz, Bezrukov:2014ipa, Bezrukov:2009db, Steingasser:2023ugv} have been accorded special significance as one can be viewed as simply an instantiation of GR with an $R^2$ correction at high energies, while the other can be viewed as the standard model Higgs with a non-minimal coupling to gravity. Furthermore, these models' predictions have been shown to coincide with a number of other models of interest \cite{Kallosh:2013tua, Kallosh:2013yoa, Ellis:2013xoa, Kallosh:2013lkr} and have consistently shown excellent compatibility with Cosmic Microwave Background (CMB) data. 

Given these prevailing attitudes, the latest results from the Atacama Cosmology Telescope (ACT) \cite{ACT:2025tim}, in combination with data from Planck \cite{Planck:2018vyg}, BICEP/Keck \cite{BICEPKeck:2022mhb}, and DESI \cite{DESI:2024mwx}, have generated a notable amount of discussion as this model now lies almost entirely outside the posterior constraints on $(n_s, r)$. Recent work includes exploring the implications that radiative corrections, reheating, or other higher order effects might have on the Starobinsky or Higgs-like models \cite{Gialamas:2025kef, Gialamas:2025ofz, Addazi:2025qra, Haque:2025uis, Zharov:2025evb, Liu:2025qca, Yin:2025rrs, Drees:2025ngb, Yogesh:2025wak, Ahmed:2025rrg, Kuralkar:2025hoz, Kuralkar:2025zxr, Modak:2025bjv}, as well as other inflation constructs \cite{Haque:2025uga, Peng:2025bws, Yi:2025dms, Maity:2025czp, Byrnes:2025kit, Kallosh:2025rni, Gao:2025onc, Bianchi:2025tyl, Heidarian:2025drk, Mondal:2025kur, Chakraborty:2025oyj, McDonald:2025tfp, Dioguardi:2025mpp, Dioguardi:2025vci, Odintsov:2025jfq, Odintsov:2025wai, Pallis:2025nrv, He:2025bli, Gao:2025viy, Frolovsky:2025iao, Zahoor:2025nuq, Ahmed:2025sfm, Lynker:2025wyc, Chen:2025qxq, Salvio:2025izr}. Here, we explore the impact that radiative corrections can have on various inflationary attractors known as $\xi$-attractors \cite{Kallosh:2013tua} and $\alpha$-attractors \cite{Kallosh:2013yoa}, the first of which can be viewed as a generalization of the Higgs-like model and the second of which can be viewed as a generalization of the Starobinsky model (for more on these attractors and their relation to each other see \cite{Odintsov:2020thl, Galante:2014ifa}). Similar to \cite{Gialamas:2025kef} which studied the impact of radiative corrections to Higgs-like inflation, we find that small, percent (and even sub-percent) level radiative correction can induce significant changes in observable predictions for these broader classes of models, and easily bring them into agreement with the latest constraints.

The paper proceeds as follows. Sec.~(\ref{sec:inflation}) discusses the $\xi$-attractor and $\alpha$-attractor models. Sec.~(\ref{sec:corrections}) follows \cite{Gialamas:2025kef, Racioppi:2018zoy, Marzola:2016xgb} in introducing a UV model agnostic way of phenomenologically parameterizing radiative corrections that could result from one loop corrections arising from the full spectrum of particle content at high energies and demonstrates that relatively small corrections can have an impact on observables. Sec.~(\ref{sec:conclusion}) concludes with a discussion of these results and their implications for assessing inflation.

\section{Models of inflation}\label{sec:inflation}

\subsection{$\xi$-attractors}

$\xi$-attractors models are scalar field models of inflation with a non-minimal coupling to gravity and are described by the following action \cite{Kallosh:2013tua},
\begin{equation}\label{Eq:xi_action}
S=\int d^4 x \sqrt{-g}\left[\frac{1}{2}\Omega(\varphi) R-\frac{1}{2}(\partial \varphi)^2-V(\varphi)\right],
\end{equation}
where $\Omega(\varphi)$ and $V(\varphi)$ have the following form,
\begin{equation}
\Omega(\varphi)=1+\xi f(\varphi), \quad V(\varphi)=\lambda f^2(\varphi).
\end{equation}
$R$ is the Ricci scalar, $\varphi$ is the scalar field, $\xi$ quantifies the scalar field's non-minimal coupling to gravity, $\lambda$ is scalar field's self-coupling, $\partial \varphi$ is the kinetic term, and $f(\varphi) = \varphi^n$. The Planck mass has been set to $M_{\mathrm{P}} =1$ throughout.

The case of $\Omega(\varphi) = 1+\xi\varphi^2$ and $V(\varphi) = \lambda\varphi^4$ (i.e.\ $f(\varphi) = \varphi^2$) corresponds to Higgs-like inflation \cite{Bezrukov:2007ep}, but this class of models is far more general and can provide viable inflationary models for many choices of $f(\varphi)$. Indeed, it can be thought of as a generalization to the historically important chaotic inflation \cite{Linde:1983gd} models to include a non-minimal coupling to gravity \cite{Kallosh:2013tua}.

The action Eq.~\eqref{Eq:xi_action} is presented in the Jordan frame which exhibits a non-minimal coupling between the scalar field and gravity. However, in order to simplify the analysis of inflationary predictions, one can perform a conformal transformation of the metric $g_{\mu \nu} \rightarrow \Omega(\varphi)^{-1} g_{\mu \nu}$ to move to the Einstein frame where the field will be minimally coupled to gravity. This transformation, combined with the following field redefinition to bring the field into its canonical form\footnote{We work in the metric formulation as opposed to the Palatini formulation, where the connection is varied independently of the metric. The Palatini formulation generally leads to different predictions for inflation observables than its metric counterpart. See \cite{Jarv:2017azx, Jarv:2020qqm, Racioppi:2019jsp, Clifton:2011jh} for some further discussion.},
\begin{equation}
\frac{d\chi}{d\varphi} = \sqrt{ \frac{1}{\Omega(\varphi)} + \frac{3}{2} \left( \frac{1}{\Omega(\varphi)} \frac{d\Omega(\varphi)}{d\varphi} \right)^2 },
\end{equation}
leads to the Einstein frame action,
\begin{equation}
S=\int d^4 x \sqrt{-g}\left(\frac{1}{2} R-\frac{1}{2}(\partial \chi)^2-U(\chi)\right),
\end{equation}
where the potential in terms of the canonically normalized field is given by
\begin{equation}
U(\chi)=\frac{ V(\varphi(\chi))}{\Omega(\varphi(\chi))^2}.
\end{equation}

\subsection{$\alpha$-attractors}

$\alpha$-attractor models \cite{Kallosh:2013hoa, Kallosh:2013yoa, Kallosh:2025ijd}, rather than arising from direct non-minimal couplings to gravity in the Jordan frame, instead arise from modifications to the kinetic term with an action such as the following,
\begin{equation}
S=\int d^4 x \sqrt{-g}\left[\frac{1}{2} R- \frac{1}{2}K(\varphi)(\partial \varphi)^2-V(\varphi)\right].
\end{equation}
Two well-known $\alpha$-attractor models are so-called ``T-models'' and ``E-models'', where T-models correspond to $K(\varphi) = 1/(1-\varphi^2/6\alpha)^2$ and E-models correspond to $K(\varphi) = 3\alpha/2 \varphi^2$ \cite{Kallosh:2025ijd}.

These can be cast into the more familiar canonical form 
\begin{equation}
S=\int d^4 x \sqrt{-g}\left(\frac{1}{2} R-\frac{1}{2}(\partial \chi)^2-U(\chi)\right),
\end{equation}
by finding the canonically normalize field $\chi$. This leads to the following analytic expressions for the potentials,
\begin{equation}\label{eq:Tmodels}
U(\chi) = V_0 \tanh^p \left( \frac{\chi}{\sqrt{6\alpha}} \right),
\end{equation}
\begin{equation}\label{eq:Emodels}
U(\chi) = V_0 \left( 1 - e^{- \sqrt{\frac{2}{3\alpha}} \chi} \right)^p,
\end{equation}
where Eq.~\eqref{eq:Tmodels} represents the T-models and Eq.~\eqref{eq:Emodels} represents the E-models. In the case of E-models, the choices $p=2$ and $\alpha=1$ correspond to the Starobinsky model. We set $p=2$ for the remainder of this paper. While the potentials are formally similar to the Starobinsky potential expressed in the Einstein frame, note here that there is no need to actually do the conformal transformation as we did with the $\xi$-attractors given that the field is not ever directly coupled to $R$ and/or never includes a higher order $R^2$ term as their origin in the simplest incarnations can be understood to come from modifications to the kinetic term (see the discussion in \cite[Sec.~10]{ Kallosh:2025ijd}). The only necessary step is to canonically normalize the field.

\subsection{Baseline observable predictions}

During inflation, quantum mechanical variations in the value of the field driving inflation will produce a spectrum of scalar and tensor perturbations in the early universe. The amplitude of scalar perturbations has been measured precisely to be $A_{\mathrm{s}} \simeq 2.1 \times 10^{-9}$ at the comoving CMB pivot scale $k_\ast = 0.05 \,\mathrm{Mpc}^{-1}$ \cite{Planck:2018jri}. The primary observable predictions used to constrain inflation models are the tensor-to-scalar ratio $r\equiv A_{\mathrm{t}}/A_{\mathrm{s}}$, which quantifies the ratio of the amplitudes of scalar and the (yet to be detected) tensor perturbations, and the scalar spectral index $n_s\left(k_*\right) \equiv 1+\left.\left(d \ln \mathcal{P}_{\mathcal{R}} / d \ln k\right)\right|_{k_*}$, which quantifies the scale dependence of the scalar power spectrum $\mathcal{P}_{\mathcal{R}}$. Finally, there is also a third observable that has generated some recent attention; the running of the spectral index $\alpha_s \equiv d n_s / d\ln k\big|_{k_*}$.
During standard scenarios where inflation is driven by the scalar field potential, the universe enters a quasi-de Sitter expansion that is well-described by the \textit{slow-roll} approximation. This slow-roll approximation \cite{Liddle:1994dx, Lidsey:1995np} is valid when $\epsilon_{\mathrm v}, \eta_{\mathrm v}, \xi_{\mathrm v}^2\ll1$, where
\begin{equation}
\begin{aligned}
\epsilon_{\mathrm v} &\equiv 
\frac{1}{2}\!\left(\frac{1}{U(\chi)} \frac{\mathrm{d} U(\chi)}{\mathrm{d} \chi}\right)^{\!2}, \quad
\eta_{\mathrm v} \equiv 
\frac{1}{U(\chi)} \frac{\mathrm{d}^2 U(\chi)}{\mathrm{d} \chi^2}, \\[4pt]
\xi_{\mathrm v}^2 &\equiv 
\frac{1}{U(\chi)^{2}}\,
\frac{\mathrm{d} U(\chi)}{\mathrm{d} \chi}\,
\frac{\mathrm{d}^{3} U(\chi)}{\mathrm{d} \chi^{3}}.
\end{aligned}
\end{equation}
Inflation ends when $\epsilon_{\mathrm v}\simeq 1$. This allows us to calculate the number of \textit{e-folds} $N(\chi)$ between the end of inflation and some earlier initial time when the fluctuations exit the horizon and freeze in the CMB,
\begin{equation}
    N(\chi) = \log \left(\frac{a_{\mathrm{e}}}{a_{\mathrm{cmb}}}\right) = \int_{\chi_{\mathrm{e}}}^{\chi_{\mathrm{cmb}}} \frac{d\chi}{\sqrt{2 \epsilon(\chi)}},
\end{equation}
where $a_{\mathrm{e}}$ is the scale factor of the universe when inflation ends and ${a_{\mathrm{cmb}}}$ is the scale factor when modes exit the horizon.
Inflation generally needs $N \simeq 55\pm 5$ e-folds to provide a viable description of the early universe. Crucially, the observable quantities $r$, $n_s$, and $\alpha_s$ can then be directly calculated from the slow-roll parameters, leading to
\begin{equation}
\begin{aligned}
    n_s - 1 &\simeq -6\,\epsilon_{\mathrm v}\!\left(\chi_{\mathrm{cmb}}\right) + 2\,\eta_{\mathrm v}\!\left(\chi_{\mathrm{cmb}}\right), \quad
    r \simeq 16\,\epsilon_{\mathrm v}\!\left(\chi_{\mathrm{cmb}}\right), \\[4pt]
    \alpha_s &\simeq
    16\,\epsilon_{\mathrm v}\!\left(\chi_{\mathrm{cmb}}\right)\eta_{\mathrm v}\!\left(\chi_{\mathrm{cmb}}\right)
    - 24\,\epsilon_{\mathrm v}^2\!\left(\chi_{\mathrm{cmb}}\right)
    - 2\,\xi_{\mathrm v}^2\!\left(\chi_{\mathrm{cmb}}\right),
\end{aligned}
\end{equation}
where quantities are evaluated when the modes exit the horizon at $\chi_{\mathrm{cmb}}$.

The $\xi$-attractors and $\alpha$-attractors have well-known predictions for $(n_s, r)$. At $\xi \ll 1$, the predictions for the $\xi$-attractors reduce to those of their minimally coupled cousins. This part of the parameter space is highly disfavored as minimally coupled polynomial chaotic models of inflation generally predict $r$ much higher than current constraints. This is not in and of itself surprising. Inflation belongs to the realm of beyond the standard model particle physics and it would not be at all surprising if e.g.\ a non-minimal coupling was needed to adequately account for observations as these are generically expected to be generated from radiative loop corrections \cite{Kaiser:2013sna}.\footnote{Interestingly, while dark energy occurs in the IR (rather than UV), recent constraints from the late time universe are likewise beginning to hint at more intricate scalar field models that feature non-minimal interactions as the simplest quintessence models do not describe current constraints on the expansion history quite as well \cite{Wolf:2024stt, Wolf:2025jed, Ye:2024ywg, Wolf:2025jlc, DESI:2025fii, Wolf:2024eph}. Of course, any non-trivial interactions will also lead to other ancillary gravitational consequences, which may bring such models into tension with other datasets. Regardless, if one attempts to model either the early or late time universe with a single scalar field, various datasets are pointing towards more intricate scalar field models.} Around $\xi \simeq \mathcal{O}(1)$ they all rapidly converge to an attractor point in $(n_s, r)$ which coincides with Starobinsky inflation, 
\begin{equation}
n_s\simeq 1- \frac{2}{N}, \quad r \simeq \frac{12}{N^2}.
\end{equation}
$\alpha$-attractors give the same predictions for $n_s$, but $\alpha$ can be chosen to reduce $r$ the arbitrarily low values,
\begin{equation}
n_s=1- \frac{2}{N}, \quad r= \frac{12 \alpha}{N^2}.
\end{equation}
As we can see, $N\simeq55$ leads to $n_s\simeq 0.964$, which now lies outside the posterior constraints with a mean value of $n_s = 0.974 \pm 0.003$, indicating a $\simeq 3\sigma$ tension.\footnote{The exact number of e-folds varies depending on the model considered and the assumptions around reheating. $N\simeq55 \pm 5$ represents a shorthand, model agnostic approximation that generally captures many of the most common models of inflation \cite{Liddle:2003as}. Even given the reheating analysis of \cite{Bezrukov:2008ut, Garcia-Bellido:2008ycs} which gives $N$ closer to the upper range of $N\simeq60$ for Higgs inflation still results in an approximately $\simeq 2\sigma$ tension.}

\section{Predictions Under Radiative Corrections}\label{sec:corrections}

\subsection{Modeling quantum corrections}

\begin{figure}[t]
   \centering
    {%
       \includegraphics[width=\columnwidth]{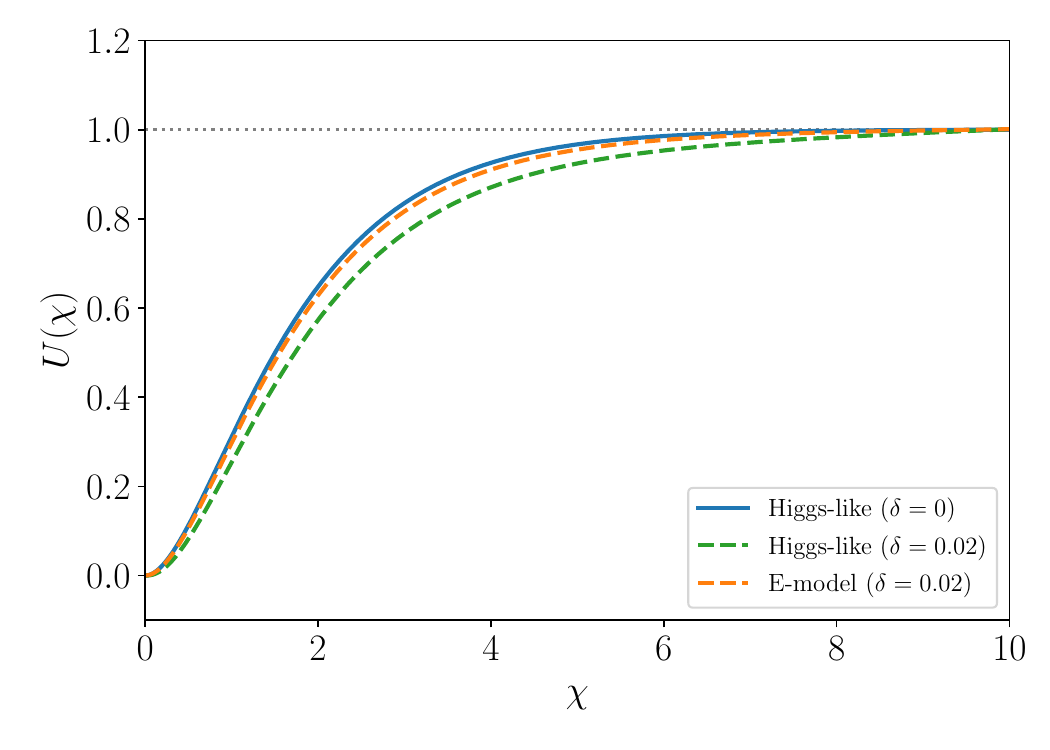}
    }
           \vskip -0.3in
    \caption{Higgs-like (with $\xi =10$) potential (solid blue line) compared with loop corrected effective potentials for the Higgs-like potential (dashed green line) and E-model $\alpha$-attractor (with $\alpha = 1$)  potential (dashed yellow line). While the Higgs-like potential and E-model potential at the chosen values for $\xi$ and $\alpha$ will coincide with the Starobinsky potential at to a very high degree of approximation at tree level, the radiatively corrected versions of these potentials differ more notably because in the Higgs-like case one must performs the conformal transformation after finding the effective potential whereas with the E-models one does not perform a conformal transformation.}\label{Fig:star_higgs_comp}
    
\end{figure}

The above are well-known inflation models and describe the dynamics of the scalar field to leading order as it is often assumed that any other fields or effects are subdominant. However, it is also generically expected that in a UV-complete theory, the full particle spectrum will induce radiative corrections to the field's self-coupling, leading to $\lambda \rightarrow \lambda (\varphi) = \lambda_{tree} + \lambda_{1-loop} (\varphi) + \cdots$.

The $\beta$-function $\beta_\lambda(\mu)=\mathrm{d} \lambda / \mathrm{d} \log \mu$ determines the running of $\lambda$, where we can expand as a Taylor series to find that the corrections take the following form,
\begin{equation}\label{eq:loop_corrections}
\lambda(\varphi, \mu_0)=\lambda\left(\mu_0\right)+\sum_{k=1}^{\infty} \frac{\beta_{k-1}}{k!} \log ^k\left(\frac{\varphi}{\mu_0}\right),
\end{equation}
where $\mu_0$ is the renormalization scale and  $\beta_k$ is $k$th derivative of the $\beta$-function at this scale. Generally speaking, Eq.~\eqref{eq:loop_corrections} is a model independent way of characterizing radiative corrections as this functional form naturally emerge from evaluating one-loop diagrams \cite{Coleman:1973jx, Racioppi:2018zoy, Gialamas:2025kef, Linde:1990flp, Martin:2013tda, Racioppi:2019jsp}. However, the exact expressions for these parameters depend on knowing the exact particle content of the UV theory, and indeed, depending on the assumptions on makes regarding UV physics, such corrections to inflationary observables could either be negligible or significant. 

To just briefly consider an example, as discussed in \cite{Fumagalli:2016sof, Burgess:2014lza, Bezrukov:2013fka, Racioppi:2019jsp}, there are numerous possible prescriptions for choosing the renormalization scale $\mu_0$. To say a bit more, $\mu$ is not truly constant and is more properly considered a function of the field $\mu(\varphi)$. For example, with one of the prescription (``prescription I'') that has been considered (chosen for its simplicity in the Einstein frame), $\mu(\varphi)$ evolves the same way as the masses of the particles being considered and merely changes the normalization of the potential without affecting its shape, meaning that inflationary observables will not change. On the other hand, another popular prescription (``prescription II'') motivated by quantization in the Jordan frame sees $\mu(\varphi)$ changing significantly during inflation; consequently, the shape of the potential is modified and radiative corrections can be significant to observable quantities (see the excellent discussion in \cite{Bezrukov:2013fka} for more details). When one chooses the first prescription (along with a few other assumptions about the theory's UV completion), it has been shown that radiative are negligible for Higgs inflation, $\xi$-attractors, and $\alpha$-attractors \cite{Fumagalli:2016sof}; yet, as also discussed there, these results do \textit{not} apply when the second renormalization prescription is chosen as these choices correspond to different UV completions and can lead to observable consequences for inflation observables \cite{Bezrukov:2013fka, Burgess:2014lza, Racioppi:2019jsp}.  Given our almost complete lack of knowledge concerning physical processes at these energy scales and the unsettled theoretical debates concerning UV completion, here we will not adopt an explicit UV completion and will pursue a different strategy.

Following \cite{Racioppi:2018zoy, Gialamas:2025kef, Artymowski:2016dlz, Marzola:2016xgb}, we work in a model agnostic manner and consider the general phenomenology that can emerge from radiative corrections to the inflation models selected here. That is, we take the first ($k=1$) term in the expansion from Eq.~\eqref{eq:loop_corrections} as the dominant correction, leading to an effective $\lambda (\varphi)$,
\begin{equation}\label{eq:jordan_loop_corrections}
\lambda(\varphi) \simeq \lambda\left[1+\delta \log \left(\frac{\varphi}{M_{\mathrm{P}}}\right)\right],
\end{equation}
where $\delta$ represents the strength of the correction, we take $\mu_0 = M_{\mathrm{P}} =1$ for convenience, and the overall magnitude of $\lambda$ can be adjusted to match the amplitude of scalar perturbations without affecting the predictions for $(n_s, r)$.

The calculation strategy of  \cite{Racioppi:2018zoy, Gialamas:2025kef} is to consider the one-loop effective potential $V(\varphi)$ by incorporating the running $\lambda(\varphi)$, and then to transform this new effective potential from the Jordan frame to the Einstein frame and canonically normalize the field to calculate inflationary observables (see \cite[fn.~2]{Racioppi:2018zoy}). We follow this same strategy for computing quantum corrections to the $\xi$-attractor models, resulting in an effective potential
\begin{equation}\label{eq:jordan_loop_corrections_potential}
U(\chi)=\frac{\lambda f(\phi(\chi))^2}{\left[1+\xi f(\phi(\chi))\right]^2}\left[1+\delta \log \left(\frac{\phi(\chi)}{M_{\mathrm{P}}}\right)\right].
\end{equation}
With the $\alpha$-attractors, there is no need to transform frames as it is not directly coupled to gravity and we can work directly with the potential expressed in terms of the canonically normalized field.
In this case, the running would then schematically written as
\begin{equation}
V_0(\chi) \simeq V_0\left[1+\delta\log \left(\frac{\chi}{M_{\mathrm{P}}}\right)\right],
\end{equation}
which can then be inserted into Eqs.~\eqref{eq:Tmodels} and \eqref{eq:Emodels} in the same manner just described.
We follow this computational strategy for both $\alpha$-attractor models (see \cite{Kallosh:2016gqp} for an example discussion of radiative corrections in $\alpha$-attractors). It is important to note that this will lead to different effective potentials for $\xi$-attractor and $\alpha$-attractor models whose uncorrected potentials coincide. For example, Fig.~\ref{Fig:star_higgs_comp} depicts the Starobinsky model (which is identical to an $\alpha$-attractor E-model with $p=2$ and $\alpha=1$) and a Higgs-like model with a quartic Jordan frame potential non-minimally coupled to gravity ($f(\varphi)= \varphi^2$ in the $\xi$-attractor language). Transforming the uncorrected Higgs-like model into the Einstein frame results in an essentially identical potential to the uncorrected E-model/Starobinsky model. However, implementing a $\delta =0.02$ correction to the Higgs-like potential and then transforming to the Einstein frame results in a different effective potential than implementing a $\delta =0.02$ correction to the E-model potential $U(\chi)$. This is because the loop correction will be affected by the conformal transformation in the Higgs-like $\xi$-attractor, whereas the loop corrections will not undergo any transformation with the $\alpha$-attractors as they already originate in the Einstein frame, despite the formal resemblances with other similar models that are conceived of as non-minimally coupled to gravity.\footnote{See \cite{Faraoni:1999hp, Chakraborty:2023kel, Clifton:2011jh, Faraoni:2006fx, Belfiglio:2024swy} for related discussion concerning the classical equivalence and possible quantum non-equivalence of the Jordan and Einstein frames.}

\subsection{Predictions for $(n_s, r)$}

\begin{figure*}[t]
  \centering
  \includegraphics[width=0.48\textwidth]{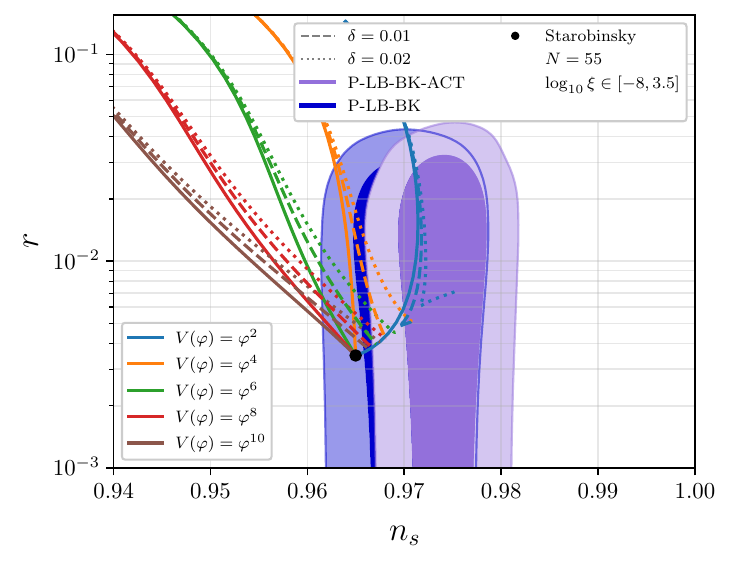}
  \hfill
  \includegraphics[width=0.48\textwidth]{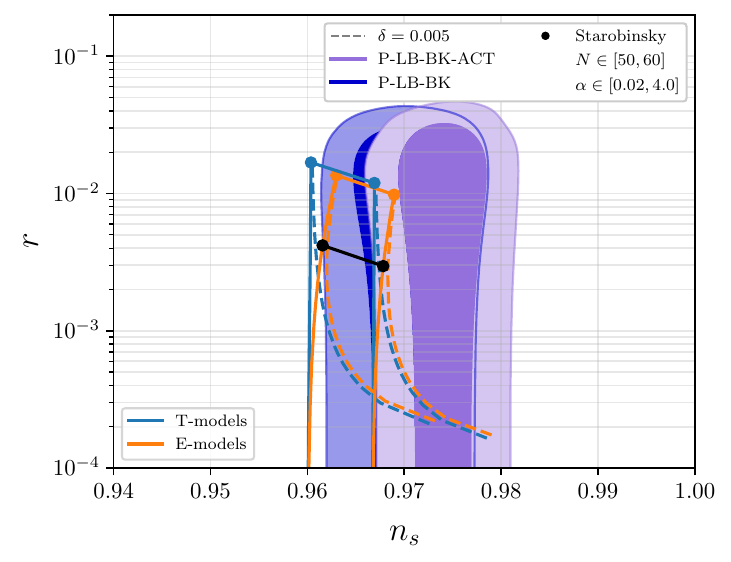}
  \caption{Comparison of $n_s$ and $r$ predictions for $\xi$-attractors (top left) and $\alpha$-attractor T and E-models (top right). Here we see that percent level quantum corrections can move the $\xi$-attractors off of their strong coupling attractor and comfortably within the latest ACT posteriors for the central $N=55$ number of e-foldings where we have considered values of the non-minimal coupling parameter $\log_{10} \xi \in [-8,\ 3.5]$. Additionally, we see that sub-percent level quantum corrections can induce fairly substantial shifts in the predictions for $\alpha$-attractors at small $\alpha$, where we have included the predictions spanning $N \in [50, 60]$ for both T and E-models considering values of $\alpha \in [.02, 4.0]$.}
  \label{Fig:xi_T_ns_r}
\end{figure*}

\begin{figure*}[t]
   \centering
    {%
       \includegraphics[width=\columnwidth]{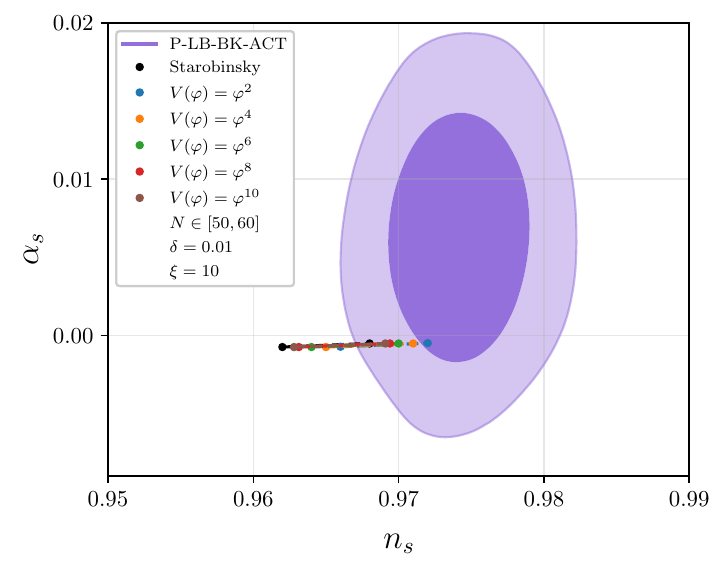}
       \includegraphics[width=\columnwidth]{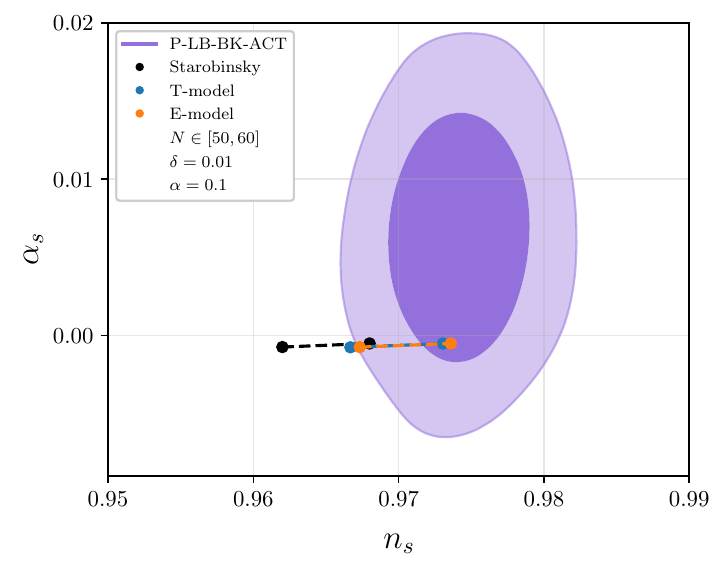}
    }
           \vskip -0.3in
    \caption{The results for $\alpha_s$ and $n_s$ (bottom) for $\xi$-attractors and $\alpha$-attractors for representative models from these classes of models ($\xi = 10$ and $\alpha = 0.1$) with a percent level correction spanning $N \in [50, 60]$. We see $n_s$ shift as expected, but the radiative corrections do not induce a notable change in $\alpha_s$.}\label{Fig:alpha_s_ns}
    
\end{figure*}

The latest results from ACT combine data from a number of probes to produce constraints on ($n_s, r$). We use P-LB-BK-ACT and the P-LB-BK combinations defined in \cite{ACT:2025tim} in order to see the difference in observables that results from including ACT. As one can see from Fig.~\ref{Fig:xi_T_ns_r}, the inclusion of ACT produces a notable shift to the right for the $n_s$ values.\footnote{In addition to these results from ACT, there has also been some recent literature on the impact that the resolution of the Hubble tension can have on the inferred values of $n_s$, which can potentially produce an even more dramatic shift towards $n_s \rightarrow1$ \cite{Wang:2024tjd, Ye:2022efx, Giare:2023wzl, Giare:2024akf}.}

Recently, \cite{Gialamas:2025kef} showed that incorporating loop corrections of the type given by Eq.~\eqref{eq:jordan_loop_corrections} into non-minimally coupled Higgs-like inflation can bring the model into better agreement with the latest results from ACT. In the case of metric variation, the authors found that corrections of strength $\delta \simeq \mathcal{O}(.01)$ can move the model from outside the data posteriors into the 1$\sigma$ region. Here we generalize these results and demonstrate that this equally applies to many models within the $\xi$-attractor family, examining $f(\varphi) = \varphi^n$ for $n=[1,2,3,4,5]$ and $\log_{10} \xi \in [-8,\ 3.5]$.\footnote{One might wonder if considering such large values of the non-minimal coupling parameter $\xi$ will cause these models to run into the unitarity problem \cite{Barbon:2009ya, Burgess:2009ea}. Here, we consider a similar range of $\xi$ as was done in the original $\xi$-attractor paper \cite{Kallosh:2013tua}. There, the authors convincingly argue that this is not a concern here because inflation in all of these models occurs at large field values ($\varphi \geq \mathcal{O}(1)$), which are several orders of magnitude larger than field values for which the problem was established.} 

The $n=1$ model is the familiar quadratic model of inflation $V(\varphi) = m^2\varphi^2$ but with a non-minimal coupling. As highlighted in \cite{Kallosh:2025rni}, is actually very compatible with the ACT data before it reaches its attractor point. The $n=2$ represents Higgs-like inflation with a quartic potential, and larger $n$ represent higher order polynomial potentials. As mentioned earlier, at $\xi \simeq \mathcal{O}(1)$ these models all converge on the predictions for the Starobinsky model, meaning that for $N=55$ e-foldings they all lie outside the latest ACT posteriors. However, with the exception of the $n=1$ model, they all approach the attractor from the left which means that at $\xi \ll 1$ they are all strongly disfavoured. 

Fig.~\ref{Fig:xi_T_ns_r} (left panel) shows that introducing a loop correction of $\mathcal{O}(10^{-2})$ (chosen to be $\delta=0.01$ and $\delta=0.02$) shifts every model to the right in $(n_s, r)$ and in better agreement with ACT, such that the $N=55$ prediction for every model considered now is compatible with either the 1$\sigma$ or 2$\sigma$ regions. This restores the compatibility of this family of inflation models at strong coupling; however, one can also see that the Starobinsky attractor has been broken with the introduction of the effective potential as the higher order polynomials are more stable against the shifts in $n_s$ and $r$ induced by loop corrections. Further increasing the size of $\delta$ then shifts the predictions for these parameters to the upper right in the $(n_s, r)$ plane, with the higher order polynomials lagging behind the lower order polynomials.

\begin{figure}[t]
   \centering
    {%
       \includegraphics[width=\columnwidth]{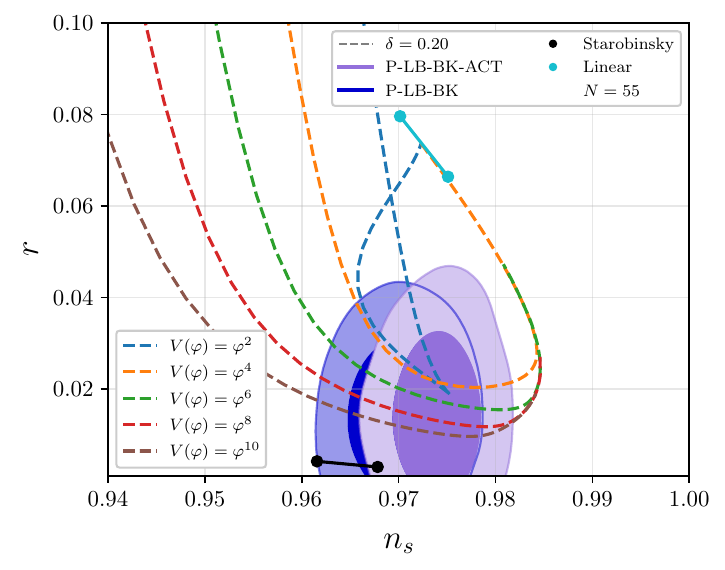}
    }
           \vskip -0.3in
    \caption{As the loop correction term $\delta$ is made larger, the entire family of $\xi$-attractors approaches a new attractor given by linear inflation. See also \cite{Artymowski:2016dlz}.}\label{Fig:linear_attractors}
    
\end{figure}

As shown by \cite{Racioppi:2018zoy, Gialamas:2025kef} for Higgs-like inflation in both the Palatini and metric variations, as $\delta$ is increased, the model begins to approach a new attractor of linear inflation $V(\varphi) \propto \varphi$. The same holds for the rest of the $\xi$-attractors considered here (see also \cite{Artymowski:2016dlz, Racioppi:2018zoy} for discussion of the linear attractor in this case). Choosing $\delta=0.2$ and $\log_{10} \xi \in [-8, 5]$, one can see in Fig.~\ref{Fig:linear_attractors} that they all begin to approach the predictions for linear inflation.
As before, the higher polynomials approach the attractor more slowly in both $\delta$ and $\xi$ as their potentials are more robust against quantum corrections.

Coming now to the $\alpha$-attractor models, they behave similarly to the $\xi$-attractor models with respect to the changes in their observables when $\alpha \simeq 1$. Although not exactly equivalent due to the aforementioned difference resulting from the conformal transformation that is performed in the case of the $\xi$-attractors, percent level loop corrections can noticeably shift their predictions to be within ACT posteriors. This is not surprising as within this regime the $\alpha$-attractor potentials are very similar to the $\xi$-attractors and give almost identical predictions. However, as $\alpha \ll 1$, $r$ can be lowered arbitrarily by tuning $\alpha$ to be lower, while $n_s$ is independent of this choice. Thus, we will explore a wide range of potentials, choosing $\alpha \in [.02, 4.0]$ which roughly spans $\log_{10} r \in [-4,-2]$ and covers most of the bounds on $r$, including going an order of magnitude below CMB-S4 projected sensitivity \cite{CMB-S4:2016ple}.

\begin{figure}[t]
   \centering
    {%
       \includegraphics[width=\columnwidth]{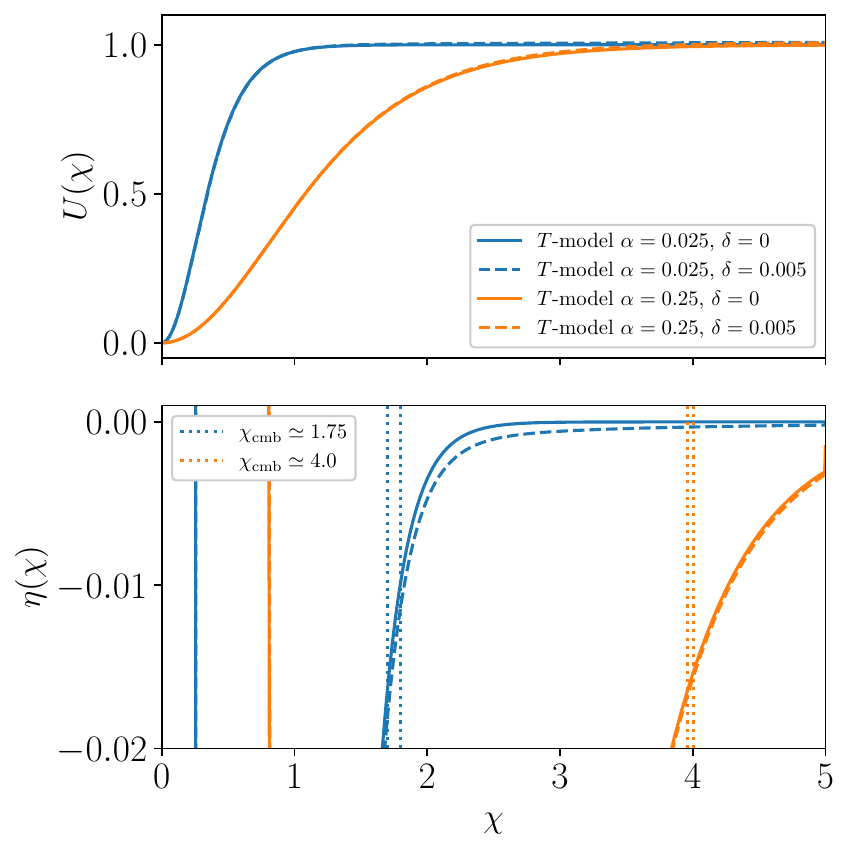}
    }
           \vskip -0.3in
    \caption{At small $\alpha$ the shift in $\chi_{\mathrm{cmb}}$ induced by the loop corrections occurs where $\eta$ is rapidly changing due to the steepness of the potential. In contrast, for larger $\alpha$, $\chi_{\mathrm{cmb}}$ occurs in a region of field space where $\eta$ is more slowly evolving, which results in a much smaller shift in the observables.}\label{Fig:T_model_comp}
    
\end{figure}

As we have already seen that percent level quantum corrections can produce a notable difference in observable predictions, it would be interesting if $\mathcal{O}(10^{-3})$ sub-percent level corrections could produce a similar effect somewhere in this parameter space. Fig.~\ref{Fig:xi_T_ns_r} (right panel) depicts the predictions for T and E-models alongside their loop-corrected effective potentials for $\delta = 0.005$. Around $\alpha \simeq 1$ we see very little departure from the predictions of the potentials without corrections, which is not surprising considering that percent level corrections were required to generate significant movement towards the central regions of the posteriors in $\xi$-attractor predictions. However, as $\alpha$ (and $r$) decrease, we begin to see far more significant departures from the uncorrected predictions. Around $\alpha \simeq 0.1$, the observables for $(n_s, r)$ evolve rapidly towards the regions favoured by the data posteriors. 

The reason for this is as follows. As $\alpha$ decreases, the potential's plateau is flatter across a larger range of field values when compared with a larger $\alpha$ model, before the field then rapidly rolls down a much steeper hill. Consequently, the region around where the modes exit the horizon will be in a regime of fields values for which the potential (and crucially its derivatives) are undergoing a sharper evolution. Thus, even a small correction which induces a slight change in the field value where the scalar modes exist the horizon can potentially change $r$ and $n_s$ significantly. This is depicted for a T-model in Fig.~\ref{Fig:T_model_comp}, where one can see that the shift in $\chi_{\mathrm{cmb}}$ induced by the loop corrections occurs where $\eta$ (the dominant contribution to $n_s$) is rapidly changing. In contrast, for larger $\alpha$, $\chi_{\mathrm{cmb}}$ occurs in a region of field space where $\eta$ is more slowly evolving, meaning that the corresponding correction to $n_s$ will be less significant.

While radiative corrections can clearly induce a notable shift in $n_s$, as also displayed in Fig.~\ref{Fig:alpha_s_ns}, introducing radiative corrections does not induce any notable changes in the values of $\alpha_s$ as these values remain small, negative, and right in line with the magnitudes expected from the standard Starobinsky model.

\subsection{Toy model with matter fields}
So far, we have calculated everything at the phenomenological level, assuming radiative corrections from unspecified field content of magnitude $\delta$. Of course, the exact corrections will depend on the field content of whatever beyond the standard model physics is lurking at higher energies. To illustrate how corrections like this might arise, we consider a simple toy model of a massive scalar field coupled to the inflaton. To the Lagrangian defined by Eq.~\eqref{Eq:xi_action}, we add a massive field $\sigma$ with an interaction term 
\begin{equation}\label{Eq:xi_action_interaction}
\frac{1}{\sqrt{-g}} \mathcal{L}=\frac{1}{2}\Omega(\varphi) R-\frac{1}{2}(\partial \varphi)^2-V(\varphi) - \frac{1}{2}(\partial \sigma)^2 -V_{\sigma}(\varphi, \sigma),
\end{equation}
where 
\begin{equation}
    V_{\sigma}(\varphi, \sigma) = \frac{1}{2} \left( m_0^2 + \kappa f(\varphi) \right) \sigma^2.
\end{equation}
Radiative corrections at the one-loop level from fields $i$ have the following general form \cite{Coleman:1973jx, Linde:1990flp, Senoguz:2008nok, Martin:2013tda, Enqvist:2013eua},
\begin{equation}\label{eq:radiative_corection_general}
\Delta V(\varphi)=\sum_i \frac{m_i^4(\varphi)}{64 \pi^2} \log \frac{m_i^2(\varphi)}{\mu_0^2},
\end{equation}
where $m^2 = V''$. For example, the mass is given by $m^2_{\sigma} = m_0^2 + \kappa f(\varphi)$. Assuming that the interaction term is large $\kappa f(\varphi) \gg m^2_0$, we can write Eq.~\eqref{eq:radiative_corection_general} and the corrected potential as,
\begin{equation}
V_{\text{eff}}(\varphi) \simeq \lambda f^2(\varphi) \left[ 1 + \frac{n \kappa^2}{64\pi^2 \lambda} \log \left( \frac{\varphi}{M_{\mathrm{P}}} \right) \right],
\end{equation}
We can then easily get an idea of the mass of the field required to produce the corrections seen in the left panel of Fig.~\ref{Fig:xi_T_ns_r} for the $\xi$-attractors. For instance, we can easily identify the $\delta$ from Eqs.~\eqref{eq:jordan_loop_corrections} and \eqref{eq:jordan_loop_corrections_potential} as
\begin{equation}
\delta = \frac{n \kappa^2}{64\pi^2 \lambda}.
\end{equation}
Fixing $\delta=0.01$, we can then solve for $\kappa$. We then explore two different choices for all $n$ considered in Fig.~\ref{Fig:xi_T_ns_r}: (i) for $\xi =1$ and $\lambda \simeq 10^{-10}$ (choosing $\lambda$ to match the amplitude of scalar fluctuations observed in the CMB) and (ii) for $\xi =10$ and $\lambda \simeq 10^{-8}$. Furthermore, here $\chi_{\mathrm{cmb}} \simeq 6.0, 5.5$ for the choices above. Both of these choices, for all $n$ considered here, result in nearly identical predictions for $(n_s, r)$ due to their attractor behavior (i.e.\ for each choice of $n$ at $\xi \simeq \mathcal{O}(1)$ or greater, the predictions are insensitive to increasing $\xi$), but have different implications for the simple toy model with an additional massive field.
We can then determine the effective mass $\mathcal{M}^2_{\sigma}\simeq \kappa f(\varphi)$ in order to produce $\delta=0.01$ radiative corrections for each member of the $\xi$-attractors. 

\begin{figure}[t]
   \centering
    {%
       \includegraphics[width=\columnwidth]{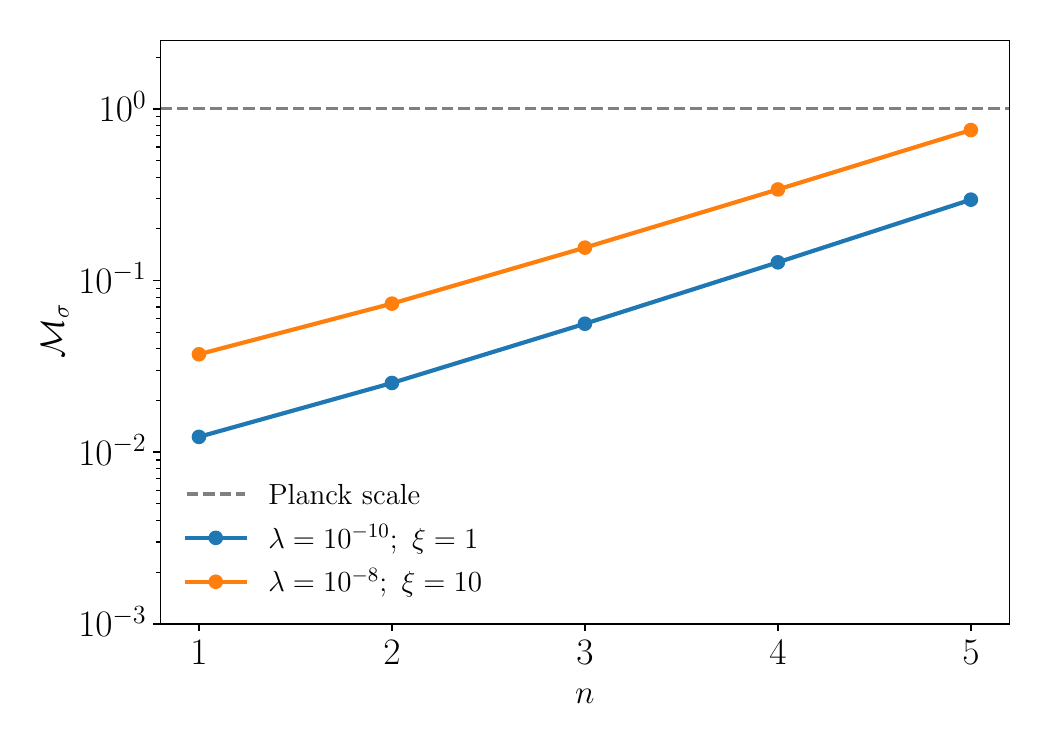}
    }
           \vskip -0.3in
    \caption{The effective mass squared of the spectator field required to produce the radiative corrections for the $\xi$-attractors in Fig.~\ref{Fig:xi_T_ns_r} for the various $n$ values considered where the potential is given by $V(\varphi) = \lambda f^2(\varphi)= \lambda \varphi^{2n}$, the non-minimal coupling is $\Omega(\varphi) = 1+\xi f(\varphi) = 1+\xi\varphi^n$, and the effective mass is $\mathcal{M}_\sigma = \sqrt{\kappa f(\varphi)}$.}\label{Fig:msplot}
    
\end{figure}

As depicted in Fig.~\ref{Fig:msplot}, this requires very heavy masses for the $\sigma$ field around or above the GUT-scale range of energies, where the mass required to produce these corrections grows approximately exponentially with $n$. Interestingly, the toy model works well at smaller non-minimal coupling strengths for all $n$ and for $n=1,2$ regardless of non-minimal coupling strength. However, the model clearly begins to breakdown at larger $\xi$ and larger $n$ as the mass required to induce the considered radiative corrections in this simple toy model approaches the Planck mass, which indicates that it cannot be trusted in this regime as we do not have a theory of quantum gravity. This is only a toy model, but it does illustrate how one might think of the origin of the phenomenological radiative corrections considered here as well as some limitations depending on the model considered. 
But of course, many other models of higher energy physics are possible. 
See also \cite{Kallosh:2016gqp} for a related discussion in the case of $\alpha$-attractors where the authors conclude that a model like this would also require GUT scale masses to generate similar radiative corrections to the ones considered in this paper.

\section{Discussion}\label{sec:conclusion}

The recent results from ACT, if they hold up to further scrutiny, at first glance seem to be indicating that several previously favoured classes of inflationary models are notably less viable than previously thought as they lie almost entirely outside the posterior constraints on $(n_s, r)$. However, the results here, as well as those of \cite{Gialamas:2025kef}, indicate that even relatively small radiative corrections can shift $n_s$ into the regions indicated by the latest ACT dataset combination. This could be interpreted favourably for these models as such corrections are seen by some as natural/expected given our knowledge of quantum field theory/standard model particle physics (assuming that the UV completion is one that would lead to non-negligible radiative corrections) and motivate the further exploration of radiative corrections in specific contexts with good theoretical motivations to gain insight into UV particle content. In particular, it may motivate finding or exploring particular UV completions that would imply corrections of these magnitudes.

On the other hand, if UV physics is such that we should expect there to be non-negligible radiative corrections to inflation, this would imply a pessimism regarding our ability to ever really say with definitive certainty what inflation predicts. After all, there would probably be far less interest in adding loop corrections, refining reheating predictions, investigating higher order effects, etc.\ if the data remained centered on the Starobinsky values. In other words, this is yet another example where higher order effects can have a non-negligible impact on observable predictions and inferences regarding model viability, such that one can effectively use such higher order effects to achieve almost any desired value to fit within observable constraints (another recent example includes hilltop models where it has been shown that higher order terms that stabilize the potential can likewise have a dramatic impact on what were previously considered to be definitive predictions for the models in question \cite{Wolf:2024lbf, Kallosh:2019jnl, Stein:2022cpk}). Amidst such uncertainty, what can we really say about inflation given that cosmological data seems be to unable to discriminate amongst the multitude of options regarding the nature of inflation, an uncertainty which also persists regarding the other two big open questions in cosmology (dark matter and dark energy \cite{Ferreira:2025fpn, Wolf:2023uno})?

First, as has been the case for a while, it appears that non-minimal couplings (or non-trivial modifications to the kinetic sector as in the case of $\alpha$-attractors) are becoming increasingly unavoidable as they play an important role in allowing inflationary models to satisfy bounds on $r$ \cite{Tenkanen:2017jih, Bostan:2018evz, Kaiser:2013sna, Kallosh:2013tua,  Reyimuaji:2020goi, Martin:2013nzq, Kaiser:2013sna, Pallis:2010wt}. And furthermore, they are widely expected to be present at these energies. Interestingly, dark energy research concerning the current epoch of accelerating expansion seems to also be following this trend of pointing towards ever more intricate scalar field constructs \cite{Ye:2024ywg, Wolf:2025jed}. So, all things considered, observations seem to be consistently pushing us towards more complicated sections of the single-field model space.

Second, there are some signals that should shed light on the viability of the single-field paradigm. For example, single-field models quite generically predict a small negative running of the spectral index $\alpha_s \equiv d n_s/d \ln k$, where the value lies between $-1.8 \times 10^{-3}<\alpha_{\mathrm{S}}<-9.1 \times 10^{-5}$ \cite{Martin:2024nlo}. Radiative corrections do not spoil this for any of the models considered here, so they still are within the scope of this generic prediction. The recent ACT results have also notably shifted the constraints on $\alpha_s$ towards the positive direction (see \cite[Fig.~4]{ACT:2025tim}). While single field inflation models like those considered here are still compatible with the constraints, in addition to the rightward shift in $n_s$, this may be another hint that some of the historically more favoured models of inflation may be in tension with new data. If constraints on $\alpha_s$ continue to migrate towards the positive direction, this may necessitate moving away from the class of simple single-field inflation altogether. Similarly, investigations relating to generating particular features in the primordial power spectrum, producing primordial black holes, and certain signals in the stochastic gravitational wave background also might have something to say regarding the ultimate viability of the more basic single-field inflation paradigm when compared with more intricate constructs within single-field modeling, multi-field models, or even more exotic alternatives \cite{Qin:2023lgo, Inomata:2021uqj, Braglia:2020fms, Lorenzoni:2025gni, Mishra:2019pzq, Garcia-Bellido:2017mdw, Wands:2007bd, Vagnozzi:2022qmc}.

\section*{Acknowledgements}

I am very grateful to David Kaiser, Pedro Ferreira, Ioannis Gialamas, Andrei Linde, Orlando Luongo, Sunny Vagnozzi, and Antonio Delgado for illuminating discussions. I am also grateful for financial support from St.~Cross College, University of Oxford.



\newpage
\bibliography{refs}

\end{document}